# Enhancing Web Application Firewalls with Machine Learning for SQL Injection Detection

**Lilliane Linnet Musoke**
Department of Computer Science
University of Reading, UK
lylliamusoke@gmail.com

**Prof. Atta Badii**
Department of Computer Science
University of Reading, UK
atta.badii@reading.ac.uk

**Ahmed Ashlam**
Department of Computer Science
University of Reading, UK
a.ashlam@pgr.reading.ac.uk

**Abstract**

Detecting SQL Injection (SQLi) attacks ranks among the most critical challenges in web application security. This research conducted a systematic literature review to identify the research gaps in this domain and responsively designed and optimised a DistilBERT-Stacked Ensemble pipeline to improve detection efficiency and robustness while reducing false-positive and false-negative rates. Comprehensive pre-processing and tokenisation were performed, DistilBERT embeddings were extracted, and machine-learning and ensemble classifiers were trained and ranked on accuracy, precision, recall and F1-score. The three best performers (Logistic Regression, XGBoost and SVM) were combined through a neural meta-learner to form a stacked ensemble. The ensemble was hardened with adversarial examples generated by the Fast Gradient Sign Method (FGSM) and tuned with Optuna. The optimised ensemble achieved 99.81% across all reported metrics, closely comparable to the strongest single model (DistilBERT-SVM, 99.82%). On the evaluation platform used in this study (Section 3.8), the ensemble classified the full test set in 0.0136s against 1.896s for DistilBERT-SVM, an approximately 140-fold reduction in measured inference latency, while retaining 99.77% accuracy under a single-step FGSM attack. The contribution is the design and validation of a SQLi detector performing with state-of-the-art accuracy at real-time speed and with demonstrated robustness to a single-step FGSM attack, rather than a marginal gain in accuracy. Sensitivity analysis further confirmed the stability of the model. These findings highlight the value of adversarial training and stacked meta-learning in building robust Web Application Firewalls (WAFs) for SQLi detection. For open validation, the dataset, test sets and models are made available at https://github.com/mlily2024/Final-project-SQL-injection-pipeline.

## 1. Motivation and Contribution

SQLi attacks are among the most damaging web threats: an adversary manipulates database queries to access sensitive data and compromise the underlying system (Amouei et al., 2022). As web applications grow in functionality, transaction volume and data sensitivity, the need for advanced defences becomes critical (Alsobhi & Alshareef, 2020). Web applications rely on rule-based and signature-matching systems as a first line of defence (Toprak & Yavuz, 2022), but these struggle with zero-day and obfuscated SQLi, suffer high false-positive and false-negative rates, add latency, and demand substantial compute (Lakhani et al., 2022).

Machine Learning (ML) has improved the effectiveness of Web Application Firewalls (WAFs) in detecting SQLi (Alghawazi et al., 2023): unlike static rules, ML models can learn from diverse HTTP request patterns and adapt to new threats. Deep learning and NLP models can improve detection accuracy and reduce error rates, yet challenges persist for ML-integrated WAFs (Shaheed & Kurdy, 2022): an evolving attack landscape, the accuracy/error-rate trade-off, high computational cost, interpretability, and susceptibility to adversarial attacks.

**Research gap.** Prior ML-WAF studies have in the main optimised for raw detection accuracy. They have rarely reported the joint trade-off between accuracy, inference latency and adversarial robustness, the three properties that jointly determine whether a detector is deployable in a real-time WAF.

**Contribution.** This paper proposes a DistilBERT-Stacked Ensemble that closes that gap. Contextual DistilBERT embeddings are fed into a set of ML classifiers; the top three (Logistic Regression, XGBoost, and SVM) are combined by a neural meta-learner hardened with FGSM adversarial training and tuned with Optuna. The novelty is not a raw accuracy improvement; the ensemble is closely comparable to the best single model, but the demonstration that a stacked, adversarially trained meta-learner **matches top accuracy at roughly 140× lower inference latency and with quantified adversarial robustness (99.77%)** makes it the more deployable choice for real-time WAFs.

## 2. The DistilBERT-Stacked Ensemble

The model combines Logistic Regression, SVM and XGBoost with DistilBERT embeddings. DistilBERT, a smaller, faster distillation of BERT (Sanh et al., 2019), processes raw text queries into contextualised embeddings that capture their semantic meaning. These embeddings serve as rich input features for the ML and ensemble classifiers, each of which generates predictions from its own architecture. The top three performers, selected by cross-validated accuracy, precision, recall, and F1-score, are stacked via a meta-learner that uses their class probabilities as meta-features.

$$y_{pred} = \propto_{LR} h_{LR}\big(E(X)\big) + \propto_{SVM} h_{SVM}\big(E(X)\big) + \propto_{XGB} h_{XGB}\big(E(X)\big) \quad (1)$$

$$where\ \propto_{LR}, \propto_{SVM}, and\ \propto_{XGB}\ are\ the\ weights\ learned\ by\ the\ \text{DistilBERT} - \text{Stacked Ensemble}$$

The meta-learner is a compact feed-forward neural network: an input layer that receives the stacked base-model probabilities, one hidden layer with ReLU activation to capture non-linear interactions, a dropout layer for regularisation, and a two-unit output layer giving the final class probabilities. It is trained with cross-entropy loss and the Adam optimiser; FGSM adversarial examples are injected during training to strengthen resilience against evasion, and Optuna tunes the architecture and training hyperparameters (Section 3.5). The model is ranked on performance, latency and sensitivity analysis.

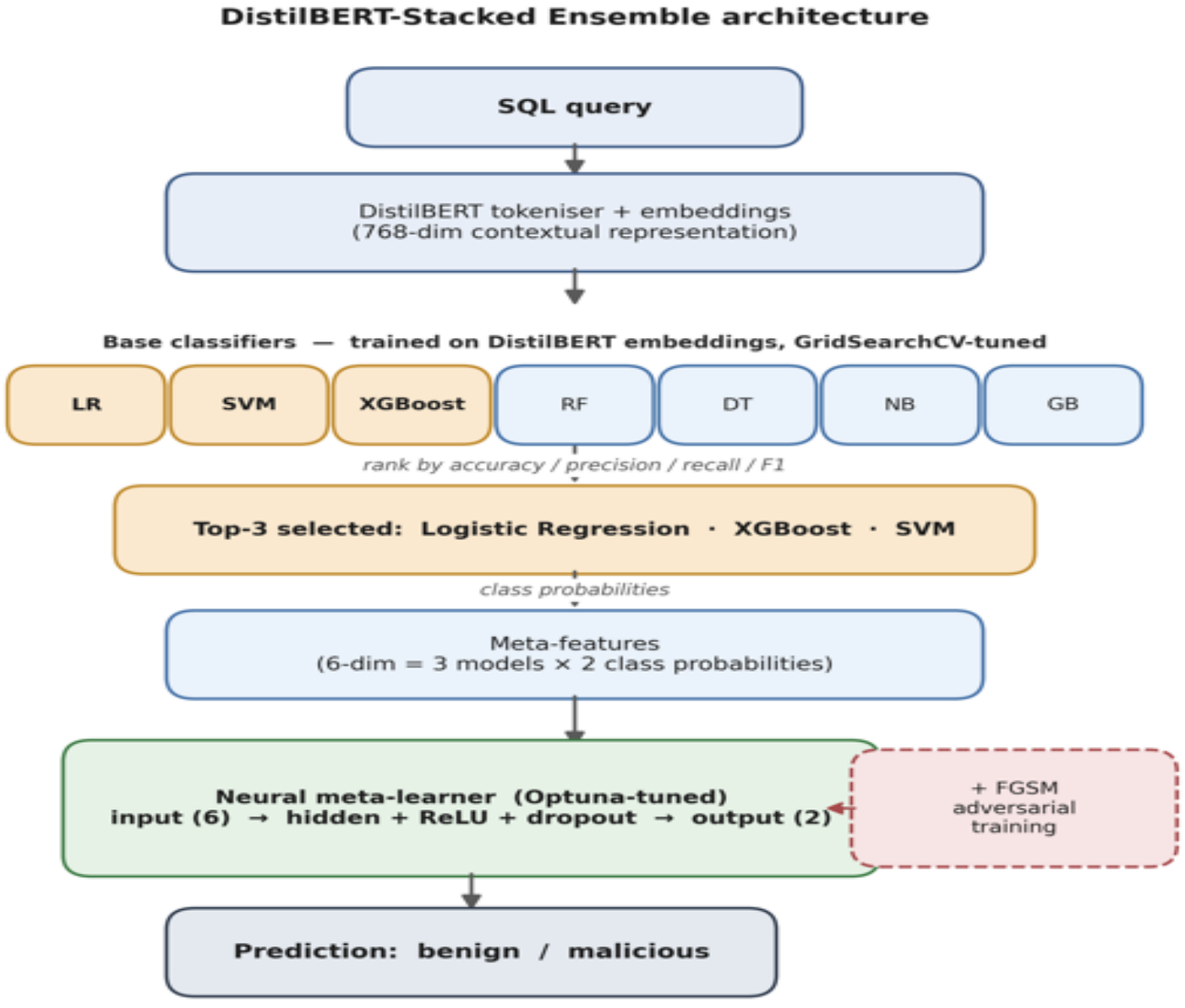

***Figure 1:*** *DistilBERT-Stacked Ensemble architecture*

***Figure 1.*** Architecture of the DistilBERT-Stacked Ensemble. Raw SQL queries are encoded into 768-dimensional DistilBERT embeddings and classified by seven base models (six tuned with GridSearchCV); the top three (Logistic Regression, XGBoost and SVM) pass their class probabilities as six meta-features to an Optuna-tuned neural meta-learner hardened with FGSM adversarial training, which outputs the benign/malicious prediction.

### 2.1 Machine Learning and Ensemble Models

A range of conventional ML and ensemble models was assessed to benchmark the proposed hybrid ***(Table 1)***. The selection reflects complementary strengths: Random Forest and Decision Trees (Suthaharan, 2016) offer robustness and interpretability; SVM (Kecman, 2005) provides strong classification in high-dimensional spaces; XGBoost (Chen and Guestrin, 2016) and Gradient Boosting (Friedman, 2001) add powerful ensemble techniques; and Logistic Regression (Hosmer and Lemeshow, 2000) and Naïve Bayes (Ting et al., 2011) offer simplicity and interpretability.

**Table 1: Machine-learning and ensemble models benchmarked.**

| Model | Description |
|---|---|
| Logistic Regression | A linear model for binary classification that estimates the probability of an input belonging to a class. Its simplicity and interpretability make it easy to understand how each feature affects the prediction (Hosmer & Lemeshow, 2000). |
| Decision Tree | A non-linear model that splits the data into subsets by feature value to form a tree: every leaf node is a class label and every internal node a decision on a feature (Suthaharan, 2016). Used for its ability to capture intricate patterns and give clear decision rules for SQLi detection. |
| Random Forest | An ensemble that builds many decision trees and combines them to achieve more reliable, accurate predictions; each tree is trained on a random subset of the data (Ho, 1995; Breiman, 2001). Reduces overfitting and variance. |
| SVM | Finds the optimal hyperplane separating benign and malicious queries. Using the kernel trick, it learns complex, non-linear relationships and performs well in high-dimensional spaces, for both linear and non-linear classification (Kecman, 2005). |
| Naïve Bayes | A probabilistic classifier that assumes feature independence under Bayes' theorem. It performs well despite its simplicity, particularly with text data, making it suitable for text classification tasks such as SQLi detection (Ting et al., 2011). |
| XGBoost | A performant, efficient implementation of gradient boosting that handles class imbalance and detects intricate data patterns (Chen & Guestrin, 2016). |
| Gradient Boosting | Combines multiple weak learners into a strong learner, learning complex relationships and |

| Model | Description |
|---|---|
| | exposing feature importance (Friedman, 2001). SQLi attacks involve complex patterns, which Gradient Boosting captures effectively, and it scales with increasing data size. |

## 3. Implementation

### 3.1 Dataset

The Kaggle SQL-injection dataset was used, containing 30,919 instances labelled malicious or benign over two fields (Query and Label). The data are imbalanced ***(Figure 2)***: 63.2% benign and 36.8% SQLi, which risks biasing models toward the majority class. For the conventional ML models the minority class was oversampled with SMOTE; for the DistilBERT-Stacked Ensemble, class weights were computed instead so that both classes carry equal weight. Balancing improved the ability of the model to detect the minority (attack) class, which matters most for web application security.

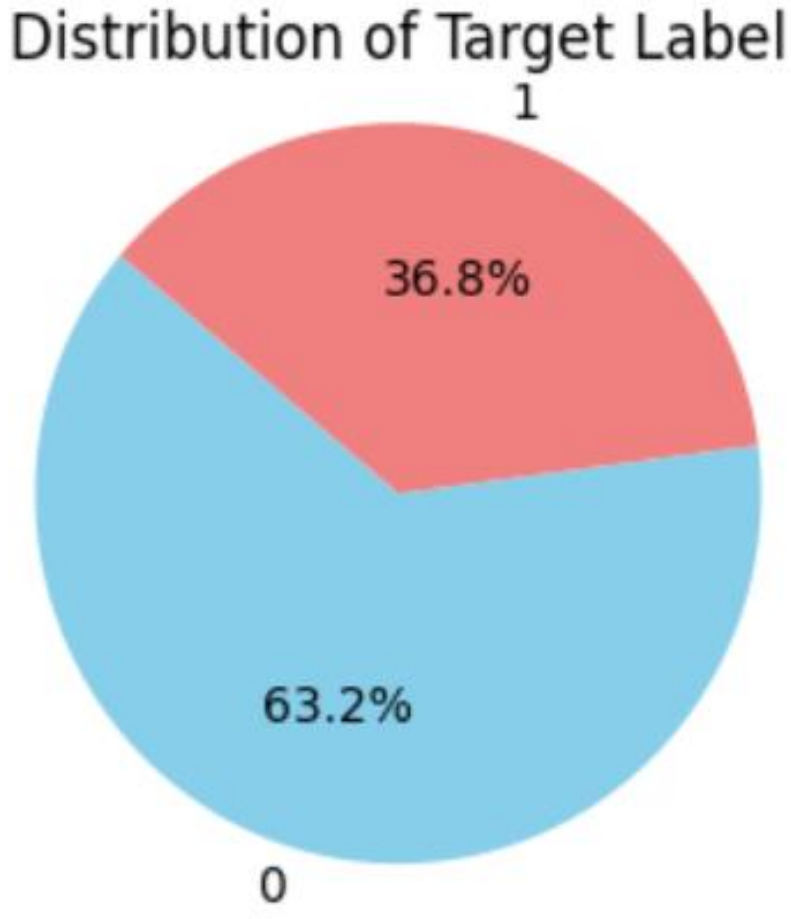

***Figure 2:*** *Distribution of Target Variable*

### 3.2 Exploratory Data Analysis and Feature Engineering

Queries were lower-cased and stripped of leading/trailing whitespace to remove case and spacing inconsistencies; no missing values were present. A Query_Length feature (word count) was added to characterise the query distribution. The 99th percentile of query length was 79 words; no rows were removed, to retain all query variations including outliers. Labels were encoded with the scikit-learn LabelEncoder, and the data were split 80/20 into training and test sets.

The correlation analysis ***(Figure 3)*** is reported over the numeric fields only Query_Length and Label. Query_Length and Label show a moderate positive correlation (0.50), indicating that longer queries carry some signal of maliciousness, consistent with attack payloads tending to be longer and more complex. (The raw Query string is not included in the correlation: it is a free text field with no meaningful numeric ordering, so any correlation computed on an integer encoding of it would not be interpretable. Its predictive value is captured instead through the DistilBERT embeddings.)

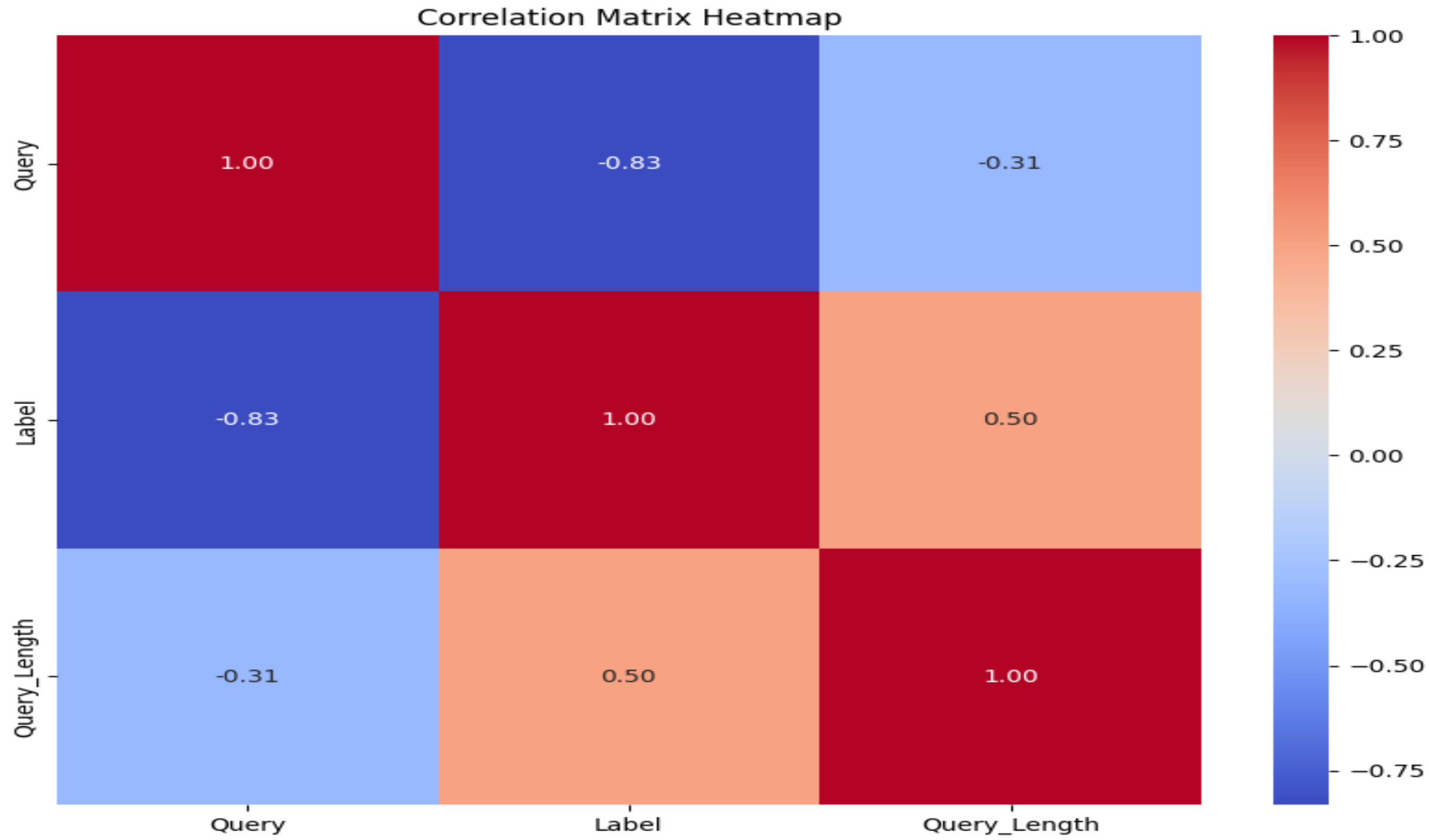

***Figure 3:*** *Correlation Matrix Heatmap*

### 3.3 Tokenisation and DistilBERT Embeddings

Queries were tokenised using the DistilBERT tokeniser, with padding and truncation to a consistent length to ensure uniform embedding computation. DistilBERT then produced multidimensional contextual embeddings that encode the syntactic and semantic details of each query. These embeddings were used as the input features for all ML and ensemble classifiers.

### 3.4 Classifier Training with DistilBERT Embeddings

Using the DistilBERT embeddings as input, the ML and ensemble classifiers were trained with GridSearchCV (5-fold, scoring on accuracy) to select the hyperparameters of each model ***(Table 2)***.

**Table 2: Hyperparameter grids searched by GridSearchCV (5-fold, accuracy).**

| Classifier | Grid searched |
|---|---|
| Random Forest | n_estimators ∈ {50, 100, 200} |
| Decision Tree | max_depth ∈ {None, 10, 20} |
| Logistic Regression | C ∈ {0.1, 1, 10} |
| SVM | C ∈ {0.1, 1, 10}, kernel ∈ {linear, rbf} |
| Gradient Boosting | n_estimators ∈ {50, 100, 200} |
| XGBoost | n_estimators ∈ {50, 100, 200} |

*Training and execution times were recorded to assess efficiency, and confusion matrices were computed for each model*.

### 3.5 Meta-Features, Meta-Learner and Optuna Optimisation

Meta-features were built from the class probabilities of the top-three base classifiers (via 5-fold cross_val_predict), giving a six-dimensional input (three models × two classes). The meta-learner architecture and training settings, after Optuna optimisation, are given in ***Table 3***.

**Table 3: DistilBERT-Stacked Ensemble meta-learner (Optuna-optimised).**

| Component | Setting |
|---|---|
| Input dimension | 6 (3 base models × 2-class probabilities) |
| Hidden layer | 170 units, ReLU |
| Dropout | 0.421 |
| Output layer | 2 units (benign / malicious) |
| Loss | Cross-entropy |
| Optimiser | Adam, learning rate 0.0083 |
| Epochs | 100 |
| Adversarial training | FGSM, perturbation budget $\varepsilon = 0.10$ |

*Optuna maximised cross-validated accuracy over the hidden size, dropout, learning rate, epoch count and FGSM ε jointly (the best trial is* ***Table 3****). Training and validation losses were logged as learning curves to check for under- or overfitting.*

### 3.6 Conventional ML and Ensemble Baselines (TF-IDF)

For comparison, the same classifiers were also trained on a conventional TF-IDF representation. Preprocessing was carried out to prepare the SQL queries for analysis by applying several steps:

- normalisation, converting all text to lowercase to ensure case-insensitivity;
- removing non-alphanumeric characters except for SQL-related characters (e.g. semicolons, quotes, and equals signs) to clean the text while preserving essential SQL syntax elements;
- tokenising the text into individual words or tokens for structural and content analysis;
- removing common English stop words while retaining important SQL keywords (e.g. select, from, where), so that SQL-specific terms critical for detecting injection patterns are not removed;
- lemmatising words to reduce them to their base or root form to improve consistency; and
- Finally, rejoining the processed tokens back into a single string for each query.

These steps cleaned and standardised the text data, improving the ability of the model to detect SQLi attacks by ensuring the text was consistent and analysable. TF-IDF was then used to convert the textual data into numerical vectors that the ML algorithms could process, assigning each word a weight according to its frequency; this improved feature extraction, reduced dimensionality, and normalised the data for more accurate and dependable models. The resulting TF-IDF features were balanced with SMOTE, and each baseline classifier was trained with its default hyperparameters, with five-fold stratified cross-validation used to report accuracy rather than the per-model GridSearchCV tuning applied to the DistilBERT-embedding classifiers (Table 2).

### 3.7 Evaluation

All models were assessed on accuracy, precision, recall and F1-score, with execution latency (wall-clock time to classify the full test set) recorded to gauge real-world viability. The statistical significance of the small accuracy difference between the stacked ensemble and the strongest single model was assessed with McNemar's exact test on their paired test predictions. A sensitivity analysis was performed on the DistilBERT-Stacked Ensemble by perturbing input features and measuring the average change in predictions, yielding per-feature sensitivity scores. Adversarial accuracy under FGSM was used to quantify robustness against deliberate evasion. This attack perturbs the six-dimensional meta-feature vector (the stacked base-model probabilities) that forms the input of the meta-learner, using a single-step FGSM perturbation at $\varepsilon = 0.10$. It therefore measures robustness to a synthetic perturbation of an internal representation rather than to a query-level evasion attempt, and does not cover stronger iterative attacks such as PGD, an epsilon sweep, or black-box transferability, which are left to future work.

### 3.8 Experimental platform

All experiments were run in Google Colab Pro on an NVIDIA Tesla T4 GPU runtime. The DistilBERT embeddings (distilbert-base-uncased, via the Hugging Face Transformers library) and the neural meta-learner were computed on the GPU, while the conventional base classifiers, implemented in scikit-learn, ran on the runtime CPU. Optuna was used for hyperparameter optimisation. All training and execution times reported in this paper, including the inference latencies in Tables 4 and 5, were measured in this environment. Latency is hardware- and implementation-dependent and is not directly comparable across platforms, whereas accuracy is platform-independent.

## 4. Results

### 4.1 Base classifiers under DistilBERT embeddings

***Table 4*** reports performance, training and execution times for the ML and ensemble models using DistilBERT embeddings. DistilBERT-SVM, DistilBERT-Logistic Regression and DistilBERT-XGBoost are the strongest single models (99.82%, 99.76% and 99.74% accuracy). DistilBERT-SVM leads on every metric but has the highest execution time (1.896s), which limits its real-time use. Logistic Regression and XGBoost balance high accuracy with low latency (0.017s and 0.042s). Random Forest and Gradient Boosting are accurate (99.45% and 99.56%) but have very long training cycles (502s and 4205s). DistilBERT-Naïve Bayes is the weakest (89.76% accuracy): this drop is expected because the dense, correlated DistilBERT embedding dimensions violate the feature-independence assumption of Naïve Bayes (contrast its 98.12% under the sparse TF-IDF representation in ***Table 5***).

**Table 4: Base-model performance with DistilBERT embeddings.**

| Model | Accuracy | Precision | Recall | F1-Score | Train (s) | Exec (s) |
|---|---|---|---|---|---|---|
| DistilBERT-Random Forest | 0.9945 | 0.9945 | 0.9945 | 0.9945 | 501.95 | 0.1299 |
| DistilBERT-Decision Tree | 0.9799 | 0.9799 | 0.9799 | 0.9799 | 198.54 | 0.003915 |
| DistilBERT-Naïve Bayes | 0.8976 | 0.8992 | 0.8976 | 0.8981 | 0.273896 | 0.0680 |
| **DistilBERT-Logistic Regression** | **0.9976** | **0.9976** | **0.9976** | **0.9976** | **28.32** | **0.0168** |
| **DistilBERT-SVM** | **0.9982** | **0.9982** | **0.9982** | **0.9982** | **471.57** | **1.8964** |

| DistilBERT-Gradient Boosting | 0.9956 | 0.9956 | 0.9956 | 0.9956 | 4204.56 | 0.0390 |
|---|---|---|---|---|---|---|
| **DistilBERT-XGBoost** | **0.9974** | **0.9974** | **0.9974** | **0.9974** | **69.02** | **0.0420** |
| **DistilBERT-Stacked Ensemble (meta-learner, pre-Optuna)** | **0.9979** | **0.9979** | **0.9979** | **0.9979** | **0.4038** | **0.0587** |

*DistilBERT-SVM, DistilBERT-Logistic Regression and DistilBERT-XGBoost were chosen as the top three for the stacked ensemble.*

### 4.2 Optimised stacked ensemble

After Optuna optimisation the DistilBERT-Stacked Ensemble reached **99.81%** across all metrics (a +0.02 percentage-point gain over its pre-optimisation 99.79%) and 99.77% adversarial accuracy under single-step FGSM ($\varepsilon = 0.10$). Figure 4 shows the confusion matrix (6 false negatives, 6 false positives). ***Figure 5*** shows the training and validation losses declining steeply in the early epochs and converging to low values, indicating stable learning and good generalisation. ***Figures 6*** and *7* show the ROC and precision–recall curves, respectively, both with AUC 1.00.

Comparing the optimised ensemble with the strongest single model, DistilBERT-SVM (99.82%), the two are near-identical in accuracy (a 0.01 percentage-point difference); a McNemar's exact test on their paired test predictions found this difference to be not statistically significant ($p \approx 1.0$), confirming that the ensemble is statistically closely comparable to the strongest single model rather than merely close in point estimate. The decisive difference is latency: measured over the full 6,184-sample test set, the optimised ensemble executes in 0.01356s versus 1.896s for SVM, roughly 140× faster, as run on the computing platform in Section 3.8. Although both are sub-millisecond per query, this 140× throughput margin is what matters for a WAF that must screen every request at line rate, making the ensemble far better suited to real-time detection. Stacking also reduces the risk of overfitting for any single base model, and adversarial training gives the ensemble a quantified robustness (99.77%) that a single accuracy-optimised model does not provide.

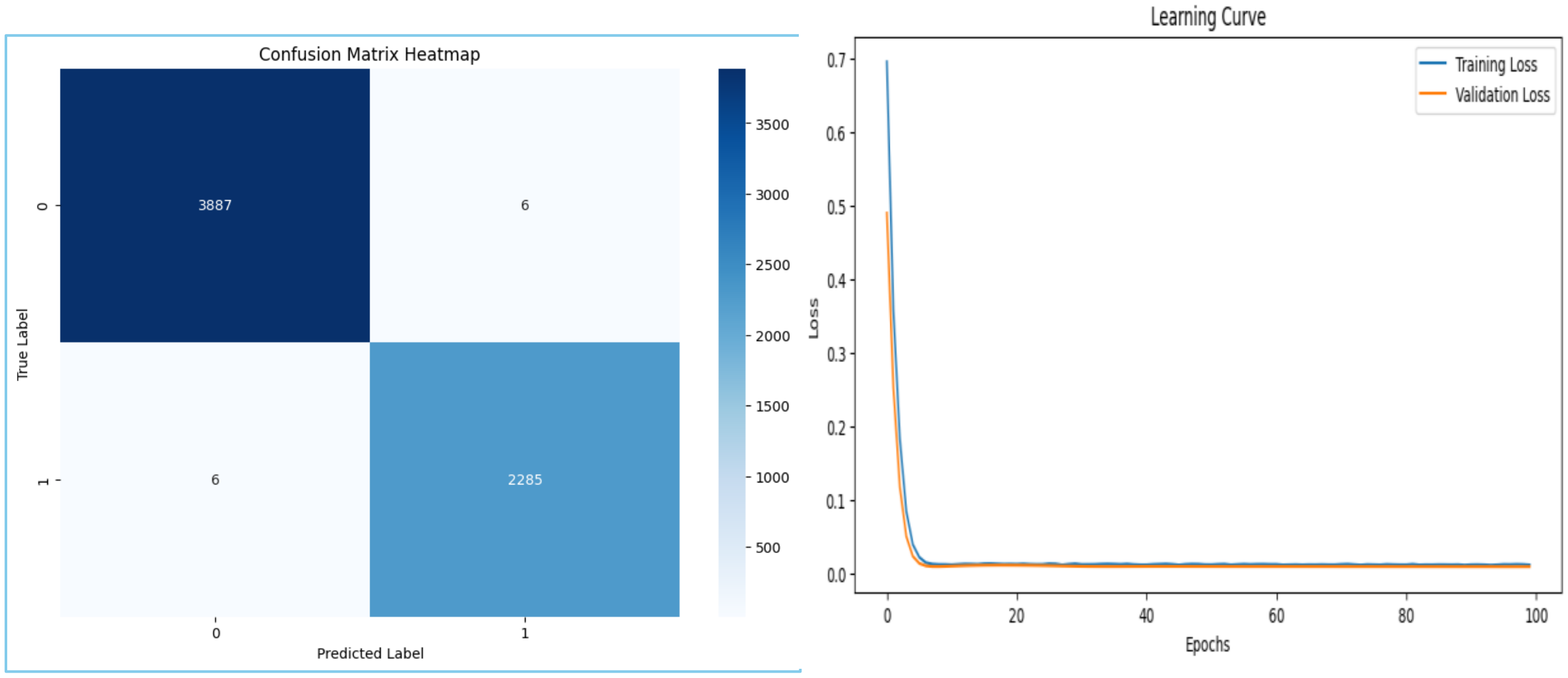

***Figure 4:*** *DistilBERT-Stacked Ensemble Confusion Matrix*

***Figure 5:*** *Learning Curve DistilBERT-Stacked Ensemble*

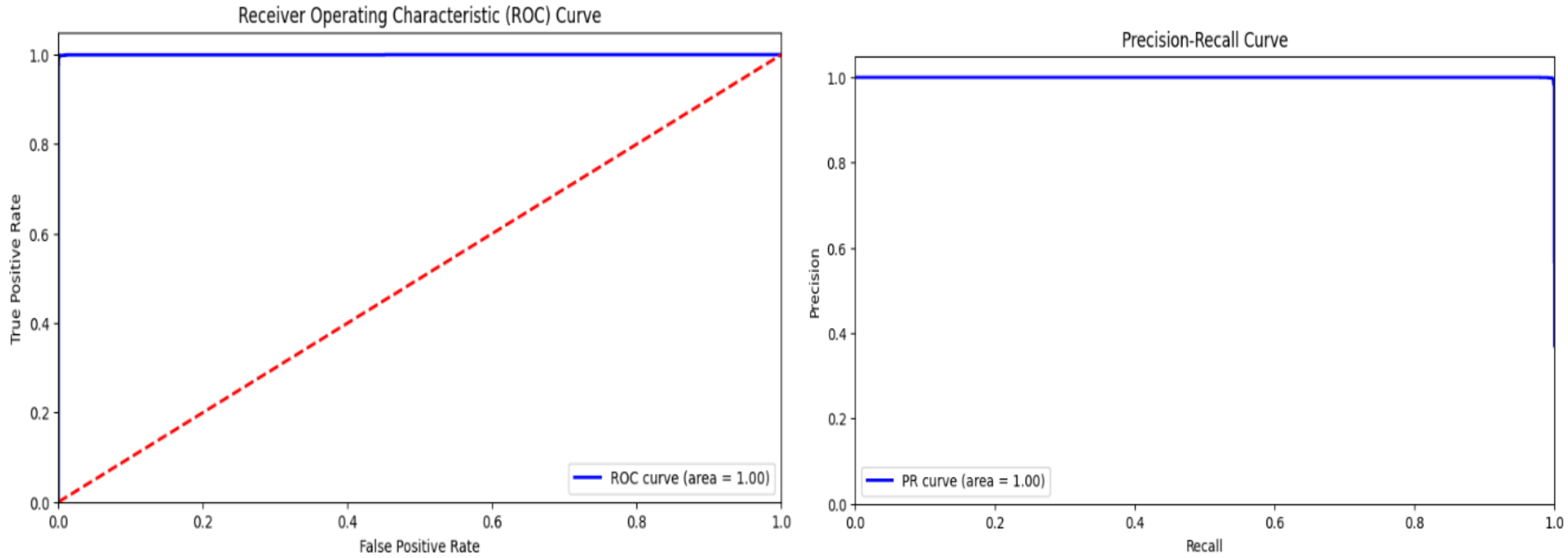

***Figure 6:*** *ROC Curve DistilBERT-Stacked Ensemble Ensemble*

***Figure 7:*** *Precision-Recall curve DistilBERT-Stacked*

## 4.3 Sensitivity analysis

The average sensitivity across features is 0.0005 ***(Figure 8)***. Features 2, 4 and 5 are the most influential; features 0, 1 and 3 less so. The low average indicates that predictions shift only moderately when inputs are perturbed, i.e. the model is stable. These scores, together with the adversarial results, indicate a pipeline that is robust to input perturbation and to deliberate evasion.

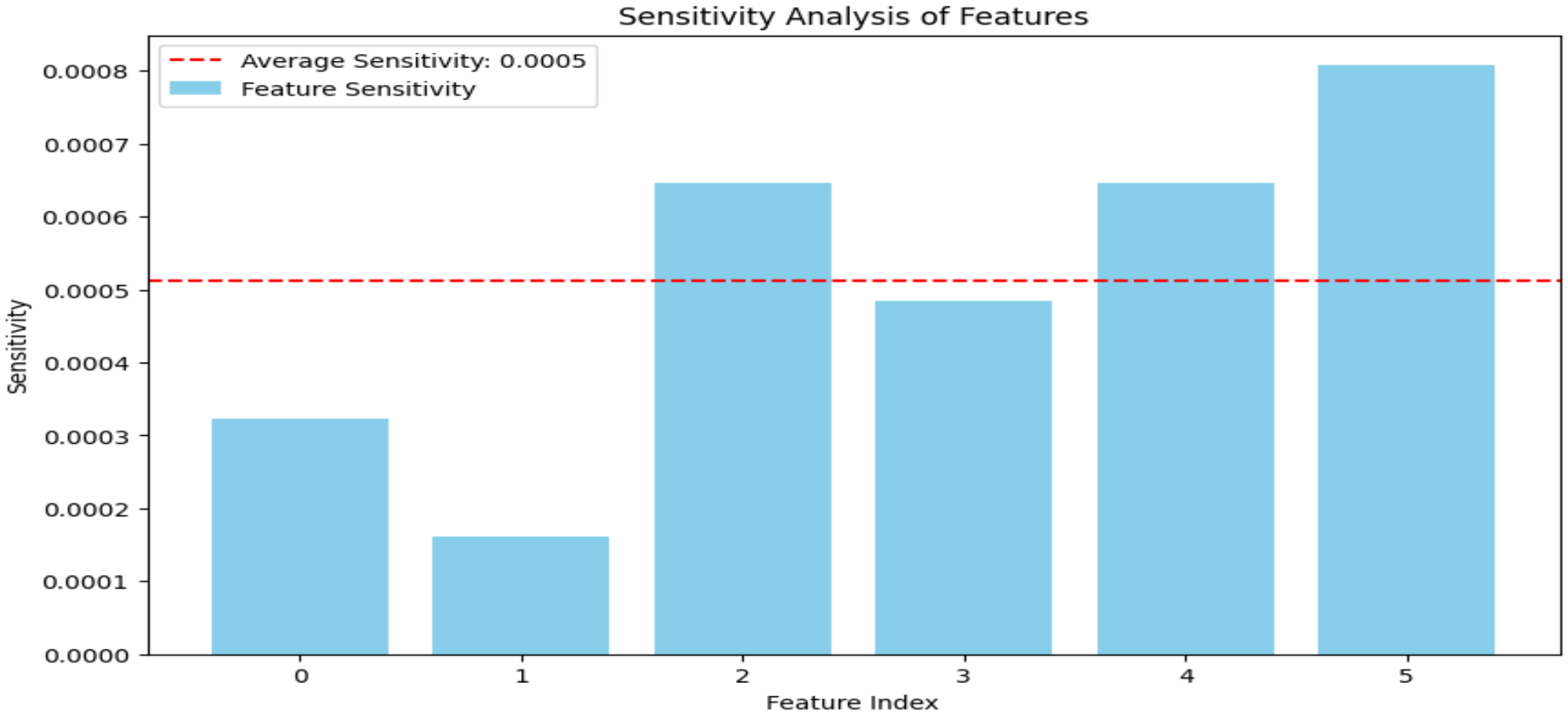

***Figure 8:*** *Sensitivity analysis DistilBERT-Stacked Ensemble*

## 4.4 Comparison with conventional ML/ensemble baselines

***Table 5*** compares the conventional TF-IDF models with the DistilBERT-Stacked Ensemble. Among the conventional models, Random Forest leads (99.47%), and Logistic Regression is the fastest. The DistilBERT-Stacked Ensemble surpasses all conventional baselines across all metrics (99.81%) while maintaining low execution time (0.01356s), confirming that DistilBERT embeddings plus stacked meta-learning improve SQLi detection in WAFs over conventional pipelines in both accuracy and efficiency. This head-to-head comparison is not fully controlled, however: the conventional baselines were balanced with SMOTE and trained with default hyperparameters, whereas the DistilBERT-Stacked Ensemble used class weighting and per-model GridSearchCV tuning, so part of the gap in Table 5 may reflect the balancing and tuning strategy rather than the embedding alone. A fully controlled comparison, with matched balancing and identical hyperparameter tuning for both representations, is left for future work.

**Table 5: Conventional (TF-IDF) models vs the DistilBERT-Stacked Ensemble.**

| Model | Accuracy | Precision | Recall | F1-Score | Exec (s) |
|---|---|---|---|---|---|
| Random Forest | 99.47% | 99.47% | 99.47% | 99.47% | 0.4624 |
| Decision Tree | 99.18% | 99.18% | 99.18% | 99.18% | 0.0045 |
| Naïve Bayes | 98.12% | 98.12% | 98.12% | 98.12% | 0.0007 |
| Logistic Regression | 98.82% | 98.82% | 98.82% | 98.82% | 0.0004 |
| SVM | 99.05% | 99.05% | 99.05% | 99.04% | 1.2386 |
| Gradient Boosting | 98.33% | 98.35% | 98.33% | 98.33% | 0.0124 |
| XGBoost | 99.39% | 99.35% | 99.39% | 99.38% | 0.0072 |
| **DistilBERT-Stacked Ensemble** | **99.81%** | **99.81%** | **99.81%** | **99.81%** | **0.01356** |

### 4.5 Query-level evasion robustness

Because the FGSM analysis above perturbs an internal representation rather than the raw query, we additionally evaluated robustness to realistic query-level evasion. Each malicious test query was rewritten with common WAF-bypass obfuscations, re-embedded with DistilBERT and scored; Table 6 reports the recall retained by the ensemble and the SVM. The detector is invariant to case and whitespace rewrites, because DistilBERT is uncased and its tokeniser normalises whitespace, and degrades only marginally under URL-encoding. Inline /**/ comments, however, are a genuine evasion vector, reducing ensemble recall to 96.4%. The SVM is marginally more robust throughout, so the ensemble provides no query-level robustness advantage over the strongest single model. This complements the meta-feature FGSM result, which measures robustness to an internal perturbation rather than to attacker-controllable query edits.

**Table 6: Query-level evasion robustness. Recall (%) on 2,291 malicious test queries after WAF-bypass obfuscation of the raw query, for the DistilBERT-Stacked Ensemble and the SVM baseline.**

| Transform | DistilBERT recall (%) | SVM recall (%) |
|---|---|---|
| **Clean (no obfuscation)** | 99.74 | 99.83 |
| **Case randomisation** | 99.74 | 99.83 |
| **Whitespace (tabs)** | 99.74 | 99.83 |
| **URL-encoding** | 99.17 | 99.61 |
| **URL-encoding + case** | 99.17 | 99.61 |
| **Inline comments (/**/)** | 96.38 | 97.21 |

## 5. Conclusions

This study designed and optimised a DistilBERT-Stacked Ensemble for detecting SQLi attacks in WAFs. The optimised ensemble demonstrated 99.81% performance capability across all metrics and 99.77% adversarial accuracy under single-step FGSM ($\varepsilon = 0.10$). A query-level evasion analysis (Section 4.5) further shows the detector resisted most query obfuscations but was evadable by inline comments, identifying a concrete hardening target. Beyond accuracy at closely comparable levels to the best performing single model, DistilBERT-SVM (99.82%), a difference that a McNemar's exact test confirms is not statistically significant, the resulting combination of the top-tier accuracy with roughly 140× lower measured inference latency (0.0136s vs 1.896s, as run on the computing platform in Section 3.8) and quantified adversarial robustness, provides advantages for deployment of the DistilBERT-Stacked Ensemble as a detector for real-time WAFs. The results underscore the value of adversarial training and stacked meta-learning for web-applications security and provide a practical framework for robust real-time SQLi detection capability within WAFs. For open validation, the dataset, test sets and models are made available at https://github.com/mlily2024/Final-project-SQL-injection-pipeline.

## 6. Future Work

Future work should expand and diversify the SQLi dataset to improve generalisation to novel attack vectors; explore semi-supervised or unsupervised learning to reduce the labelled-data burden; develop stronger adversarial-training methods to sustain robustness against emerging threats; validate the pipeline through dynamic, real-time WAF integration; and fine-tune the model against specific obfuscation techniques such as encryption, code and data obfuscation, and steganography.